\documentclass[reprint,aps,prb,amsmath,amssymb,twocolumn,showkeys, superscriptaddress,floatfix,showkeys]{revtex4-2}
\usepackage{graphicx}
\usepackage{dcolumn}
\usepackage{bm}
\usepackage[colorlinks = true, allcolors = blue]{hyperref}
\usepackage{verbatim}
\usepackage{soul}
\usepackage{float}
\usepackage{xcolor}
\usepackage{multirow}
\usepackage{booktabs}
\usepackage{tabularx}
\newcolumntype{Y}{>{\raggedleft\arraybackslash}X} 

\begin{document}
	\title{Anisotropic two-dimensional magnetoexciton with exact center-of-mass separation}
	
	\author{Dang-Khoa D. Le}
	\affiliation{Computational Physics Key Laboratory K002, Department of Physics,
		Ho Chi Minh City University of Education, Ho Chi Minh City 72759, Vietnam}
	
	\author{Hoang-Viet Le}
	\affiliation{Computational Physics Key Laboratory K002, Department of Physics,
		Ho Chi Minh City University of Education, Ho Chi Minh City 72759, Vietnam}
	
	\author{Dai-Nam Le}
	\affiliation{Department of Physics, University of South Florida,
		Tampa, Florida 33620, USA}
	
	\author{Duy-Anh P. Nguyen}
	\affiliation{Thu Dau Mot University,
		Phu Loi, Ho Chi Minh City 75110, Vietnam}
	
	\author{Thanh-Son Nguyen}
	\affiliation{Faculty of Fundamental Sciences,
		University of Architecture Ho Chi Minh City,
		Ho Chi Minh City 722700, Vietnam}
	
	\author{Ngoc-Tram D. Hoang}
	\affiliation{Computational Physics Key Laboratory K002, Department of Physics,
		Ho Chi Minh City University of Education, Ho Chi Minh City 72759, Vietnam}
	
	\author{Van-Hoang Le}
	\affiliation{Computational Physics Key Laboratory K002, Department of Physics,
		Ho Chi Minh City University of Education, Ho Chi Minh City 72759, Vietnam}
	
	\date{\today}

	\begin{abstract}

	Excitons in anisotropic two-dimensional (2D) materials, defined by direction-dependent effective masses, are of pronounced interest for their roles in excitonic and magneto-optical phenomena. A perpendicular magnetic field complicates the separation of center-of-mass (c.m.) and relative motions, especially when electron and hole masses are comparable. Conventional theories often employ an approximate c.m. separation using factorized wave functions, modifying magnetic Hamiltonian terms and possibly introducing inaccuracies in magnetoexciton energy predictions. This work develops an exact analytical framework for c.m. and relative motion separation in anisotropic 2D magnetoexcitons, without resorting to the stationary-c.m. approximation. Starting from the full electron-hole Hamiltonian in a homogeneous magnetic field, the formalism uses the conserved pseudomomentum to derive a relative-motion Hamiltonian, revealing new anisotropy-dependent couplings and magnetic coefficients absent in approximate models. The resulting Schrödinger equation is treated via the Feranchuk-Komarov operator method and Levi-Civita transformation, allowing non-perturbative, systematically convergent solutions. Application to monolayer black phosphorus and titanium trisulfide, both freestanding and encapsulated in hexagonal boron nitride, yields magnetoexciton energies, diamagnetic coefficients, and probability densities for the ten lowest states across considerable magnetic-field ranges. The results demonstrate the significant influence of anisotropy-dependent coupling on magnetic response in systems with strong mass anisotropy. This formalism is generalizable to other anisotropic 2D semiconductors, establishing a foundation for advanced magneto-optical studies. 
	\end{abstract}
	
	
	\maketitle
	
	\section{Introduction}
	
	The exploration of two-dimensional (2D) van der Waals materials has revealed a wide range of electronic and optical properties governed by strong Coulomb interactions and reduced dielectric screening \cite{Geim2013, Chernikov2014, Wang2018, Laturia2018,  Zhou2019, Arora2021, Meineke2024, Lian2024, Okamura2025}. In many 2D semiconductors, excitons dominate the optical response due to their large binding energies \cite{Rodin2014, Choi2015, Hill2015, Stier2018, Nguyen2019, Goryca2019, Hsu2019, chennano2019, Liu2019, Ly2023, takahashi2024, Ly2025, Lu2024}. While numerous studies have focused on isotropic systems, many low-symmetry 2D materials exhibit pronounced in-plane anisotropy, resulting in direction-dependent effective masses and strongly anisotropic excitonic properties \cite{Ferreira2017, Junior2019, Li2020, Suk2024, Dang2025, Vannucci2020, Gjerding2021}. Recent surveys indicate that effective mass anisotropy is a widespread feature in two-dimensional materials \cite{Vannucci2020, Gjerding2021}. Such anisotropy plays an important role in determining polarization-dependent absorption, carrier dynamics, and magneto-optical responses in materials such as monolayer black phosphorus (BP) and titanium trisulfide (TiS$_3$) \cite{Tran2014, Chaves2015,   Wang2015, Island2016, Villegas2016,   Paez2016, Torun2018, Donck2018, Zhang2018,   HenriquesPeres2020, Wang2021,   Wu2022, Kezerashvili2022, Kezerashvili2022b}. These unique properties make them promising for applications in electronic devices, polarization-resolved optics, and magneto-optical spectroscopy \cite{Li2014, Avsar2015, Das2021, Liu2021pers, Huang2022}. 
	
	Magnetoexciton spectroscopy provides a powerful tool to probe effective masses, screening lengths, surrounding dielectric constants, and excitonic structure in 2D semiconductors \cite{Stier2018, Nguyen2019, Goryca2019, Hsu2019, chennano2019, Liu2019, Ly2023, takahashi2024, Ly2025}. Recent high-field magneto-optical measurements in anisotropic and low-symmetry two-dimensional materials have further demonstrated the sensitivity of exciton energies and diamagnetic shifts to strong magnetic fields, highlighting the need for quantitatively reliable theoretical models \cite{Meineke2024, Okamura2025}. In isotropic systems, the exciton energy spectrum in a perpendicular magnetic field follows a well-established picture: rotational symmetry preserves magnetic quantum numbers, and the magnetic field dependence of exciton energies encodes the competition between magnetic confinement and the Coulomb attraction, described by the Rytova-Keldysh potential \cite{Rytova1967, Keldysh1979}. In anisotropic 2D materials, however, the loss of rotational symmetry can lead to hybridization between angular-momentum channels and direction-dependent diamagnetic shifts, which consequently non-trivially modify both energy spectra and wave functions. Therefore, accurate theoretical descriptions are essential for quantitatively interpreting experimental data and extracting material parameters from magnetoexciton energies.
	
	A fundamental aspect of the magnetoexciton problem is the separation of center-of-mass (c.m.) and relative motions. In homogeneous magnetic fields, these degrees of freedom are coupled, and the separation is nontrivial. For isotropic systems, rigorous treatments based on the conserved pseudomomentum provide an exact procedure for isolating relative motion \cite{Avron1978, Johnson1983,   Ruder1994, Butov2001, Ly2023b}. In anisotropic materials, where effective masses differ along principal directions, additional coupling terms arise in the Hamiltonian and modify magnetic-field–dependent contributions. In several theoretical studies of anisotropic magnetoexcitons \cite{Kezerashvili2022, Kezerashvili2022b}, the c.m. separation has been performed approximately by assuming that the total wave function factorizes into independent c.m. and relative components. This assumption is formally justified in electron–nucleus systems, where the nuclear mass dominates over the electron mass \cite{Landau1977}. This approach simplifies the calculations and enables the practical evaluation of magnetoexciton energies. However, for excitons, when the electron and hole effective masses are comparable, the magnetic-field-induced coupling between c.m. and relative coordinates may modify the structure of the effective Hamiltonian. Therefore, a careful and systematic treatment of the c.m. separation is required in order to obtain quantitatively reliable magnetoexciton energies and diamagnetic coefficients. This issue becomes particularly relevant in the interpretation of high-field magneto-optical spectra, where small deviations in the magnetic coefficients may lead to noticeable differences in the extracted material parameters.
	
	In the present work, we derive an exact analytical procedure for separating the c.m. and relative motions of an anisotropic two-dimensional exciton in a perpendicular magnetic field without invoking the stationary-c.m. approximation. Starting from the full electron–hole Hamiltonian, we employ the conserved pseudomomentum to construct the transformed relative-motion Hamiltonian. The resulting equation contains anisotropy-dependent coupling terms and modified magnetic coefficients that directly affect the magnetic contribution to the exciton binding energy. Our study considers the isotropic electron-hole interaction; however, the formalism can also be applied to materials with additional anisotropy in permittivity. In this case, the electron-hole interaction should instead be described by the anisotropic Keldysh potential \cite{Galiautdinov2019}.
	
	To obtain numerical solutions, we solve the corresponding Schrödinger equation using the Feranchuk–Komarov (FK) operator method \cite{Feranchuk1982, Feranchuk2015} combined with the application of the Levi-Civita transformation \cite{Hoang1993}. This non-perturbative approach yields systematically convergent solutions with controllable precision, allowing accurate determination of magnetoexciton energies and wave functions. We apply the formalism to monolayer BP and TiS$_3$ in both freestanding (FS) and hexagonal boron nitride (hBN)-encapsulated environments. Magnetoexciton energies, diamagnetic coefficients, and probability densities are calculated over a broad magnetic-field range for the ten lowest excitonic states, and comprehensive numerical data tables are provided. These results provide a consistent and reproducible reference for future theoretical and experimental investigations of anisotropic magnetoexcitons.
	
	The rest of the paper is organized as follows. In Sec.~\ref{sec2} we present the anisotropic electron–hole Hamiltonian and derive the exact c.m. separation and the resulting relative-motion equation. In Sec.~\ref{sec3} we outline the computational method, discuss the labeling of quantum states in anisotropic systems, and present numerical results for magnetoexciton energies and wave functions. Section~\ref{concl} summarizes the main findings and outlines possible extensions of the present formalism.
	
	\section{The Schr\"{o}dinger equation for an anisotropic 2D magnetoexciton}\label{sec2}
	\subsection{Hamiltonian of electron–hole system in a homogeneous magnetic field}
	\label{subsecIIA}
	
	We consider an exciton composed of an electron and a hole, confined in a two-dimensional plane, that interact via the Rytova-Keldysh potential \cite{Rytova1967, Keldysh1979} in the presence of a perpendicular magnetic field $\mathbf{B}$ applied perpendicular to the monolayer plane. The $x$ and $y$ directions are chosen to coincide with the principal axes of the effective mass tensor. The Hamiltonian of the system can be written as
	\begin{eqnarray}\label{eq1}
		\hat{H} &=
		\dfrac{1}{2m^e_{x}}\,\left( {\hat{p}_{x}^e}+ eA_x^e\right)^2 
		+\dfrac{1}{2m^e_{y}}\,\left( {\hat{p}_{y}^e}+ eA_y^e\right)^2 \nonumber\\
		&+\dfrac{1}{2m^h_{x}}\,\left( {\hat{p}_{x}^h}- eA_x^h\right)^2 
		+\dfrac{1}{2m^h_{y}}\,\left( {\hat{p}_{y}^h}- eA_y^h\right)^2 \nonumber\\
		&+ \,  V_{RK}(|{\mathbf{r}_e- \mathbf{r}_h}|),
	\end{eqnarray}
	where $m_{x}^e$, $m_{y}^e$, $m_{x}^h$, and $m_{y}^h$ denote the effective masses of the electron and hole
	along the $x$ and $y$ directions, respectively. Due to anisotropy, the effective masses in the two principal directions are different. The operators $\hat{p}_{x}^e$, $\hat{p}_{y}^e$, $\hat{p}_{x}^h$, $\hat{p}_{y}^h$ represent the momentum components of the electron and hole. The interaction potential $V_{RK}$ depends only on the relative coordinate $|\mathbf{r}_e- \mathbf{r}_h|$ . Here, $e$ is the magnitude of the electron charge and $B$ is the magnetic field strength. Because the electron-hole interaction is still isotropic, it is better to adopt the symmetric gauge for the magnetic vector potential, whose components are given by  $A^e_x=-By_e/2$, $A^e_y=+Bx_e/2$, $A^h_x=+By_h/2$, and $A^h_y=-Bx_h/2$.
	
	To reduce the number of variables in Eq.~\eqref{eq1} and separate collective and internal motions, we introduce relative and center-of-mass coordinates
	\begin{eqnarray}
		&x= x_e-x_h,\qquad y=y_e-y_h,\nonumber\\
		&X = \dfrac{m_{x}^e \,x_e + m_{x}^h \,x_h}{m_{x}^e + m_{x}^h}, 
		\qquad
		Y = \dfrac{m_{y}^e\, y_e + m_{y}^h\, y_h}{m_{y}^e + m_{y}^h}.
		\label{eq2}
	\end{eqnarray}
	After performing the coordinate transformation and carrying out the corresponding algebraic manipulations, the Hamiltonian expressed in the variables ($x, y, X, Y$) takes the form
	\begin{eqnarray}\label{eq3}
		\hat{H} &=& \dfrac{1}{2M_x}{{{\hat{P}}_X}}^2 + \dfrac{1}{2M_y}\hat{P}_Y^2 \nonumber\\
		&&+ \dfrac{1}{2\mu_x}\hat{p}_x^2 + \dfrac{1}{2\mu_y}\hat{p}_y^2
		+ {\hat H}_B + V_{RK} (r),
	\end{eqnarray}
	where ${\hat H}_B$ is the magnetic-field-dependent contribution, and the effective masses are defined as
	\begin{eqnarray}\label{eq4}
		M_x = m_{x}^e + m_{x}^h,\quad M_y = m_{y}^e + m_{y}^h,\nonumber\\
		\mu_x = \frac{m_{x}^e \,m_{x}^h}{m_{x}^e + m_{x}^h}, \quad 
		\mu_y = \frac{m_{y}^e\, m_{y}^h}{m_{y}^e + m_{y}^h}.
	\end{eqnarray}
	Here, $M_x$ and $M_y$ represent the c.m. masses along the two principal directions, while $\mu_x$ and $\mu_y$ are the corresponding reduced masses for the relative motion. The operators $\hat P_X$ and $\hat P_Y$ denote the c.m. momentum components, 
	and $\hat p_x$ and $\hat p_y$ represent the relative-motion momenta. The relative distance is given by $r=\sqrt{x^2+y^2}$. 
	
	The magnetic-field-dependent contribution consists of linear and quadratic terms in the magnetic field $B$ as ${\hat H}_B= {\hat H}_{\mathrm{lin}} + {\hat H}_{\mathrm{quad}}$, where 
	\begin{eqnarray}
		\hat{H}_{\mathrm{lin}} &= &
		\dfrac{eB}{2} \Bigg[  \left(\dfrac{m_{y}^e}{m_{x}^h} -\dfrac{m_{y}^h}{m_{x}^e} \right) \dfrac{y}{M_y} \hat{p}_x
		- \dfrac{y}{M_x}  \hat{P}_x - \dfrac{Y}{\mu_x}  \hat{p}_x \nonumber\\
		&&-  \left(\dfrac{m_{x}^e}{m_{y}^h} -\dfrac{m_{x}^h}{m_{y}^e}\right) \dfrac{x}{M_x} \hat{p}_y
		+ \dfrac{x}{M_y} \hat{P}_y + \dfrac{X}{\mu_y}  \hat{p}_y \Bigg]\nonumber
	\end{eqnarray}
	and
	\begin{eqnarray}
		{\hat{H}}_{\mathrm{quad}} &=&
		- \dfrac{e^2 B^2}{8} \Bigg[
		\dfrac{X^2}{\mu_y}  +  \left( \dfrac{{m_{x}^h}^2}{{m_{y}^e}^2}
		+\dfrac{{m_{x}^e}^2}{{m_{y}^h}^2} \right)\dfrac{x^2}{{M_x}^2}  \nonumber\\
		&&+ \left( \dfrac{m_{x}^h}{m_{y}^e} - \dfrac{m_{x}^e}{m_{y}^h} \right) \dfrac{2xX}{M_x} 
		+ \left( \dfrac{m_{y}^h}{m_{x}^e} - \dfrac{m_{y}^e}{m_{x}^h} \right) \dfrac{2yY}{M_y} \nonumber\\
		&& +\dfrac{Y^2}{\mu_x} 
		+  \left( \dfrac{{m_{y}^h}^2}{{m_{x}^e}^2}  + \dfrac{{m_{y}^e}^2}{{m_{x}^h}^2} \right) \dfrac{y^2}{{M_y}^2}
		\Bigg].\nonumber
	\end{eqnarray}
	
	The Hamiltonian clearly contains coupling terms between the relative and c.m. coordinates. Therefore, the variables cannot be separated in a straightforward manner, and a more careful transformation is required. This exact separation procedure will be presented in the next subsection.
	
	\subsection{Exact center-of-mass separation}
	
	To proceed with the analysis, it is necessary to separate the c.m. and relative motions from the Hamiltonian~\eqref{eq3}. In the presence of a homogeneous magnetic field, however, this separation is nontrivial because the c.m. and relative coordinates are coupled through the magnetic term $\hat H_B$. In particular, the canonical momentum associated with the c.m. motion is no longer conserved due to the vector potential.
	
	Nevertheless, the system possesses a conserved quantity known as the \emph{pseudomomentum} \cite{Avron1978, Johnson1983}, represented by an operator defined as  
	\begin{equation}\label{eq5}
		\hat{\mathbf{P}}_0 = \hat{\mathbf{P}} - \frac{1}{2} e\, \mathbf{B} \times \mathbf{r},
	\end{equation}
	where $\mathbf{P}=P_X \mathbf{i} + P_Y \mathbf{j}$ is the total canonical momentum and $\mathbf{r}= x\,\mathbf{i} + y \,\mathbf{j}$ is the relative coordinate.
	It can be verified directly that $[\hat{\mathbf{P}}_0, \hat{H}] = 0$, which implies that $\hat{\mathbf{P}}_0$ and $\hat{H}$ share a common set of eigenfunctions. We therefore seek solutions of the form 
	\begin{equation}\label{eq6}
		\Psi(\mathbf{r}, \mathbf{R}) = \exp\!\left[\frac{i}{\hbar} \left(\mathbf{K} + \tfrac{1}{2} e\, \mathbf{B} \times \mathbf{r}\right) \cdot \mathbf{R}\right] \psi(\mathbf{r}),
	\end{equation}
	where $\mathbf{R}= X\,\mathbf{i} + Y \,\mathbf{j}$, and $\mathbf{K}=K_x\, \mathbf{i} + K_y \,\mathbf{j}$ is the eigenvalue of the pseudomomentum operator. 
	Substituting Eq.~\eqref{eq6} into the Schr{\"o}dinger equation and performing the unitary transformation
	$\hat{U} = \exp\!\left[\frac{i}{\hbar} (\mathbf{K} + \tfrac{1}{2} e\, \mathbf{B} \times \mathbf{r}) \cdot \mathbf{R}\right]$, we obtain 
	\begin{equation}\label{eq7}
		\hat{U}^{-1} \hat{H} \hat{U} \psi(x,y) = E \psi(x,y).
	\end{equation}
	The relative-motion Hamiltonian is then defined as 
	\begin{equation}\label{eq8}
		\hat{H}_{rel} = \hat{U}^{-1} \hat{H}\, \hat{U} -\frac{1}{2M_x}K_x^2-\frac{1}{2M_y}K_y^2,
	\end{equation}
	where the subtracted terms correspond to the kinetic energy of the c.m. motion. 
	
	Carrying out the transformation explicitly, we obtain the relative-motion Hamiltonian in the form
	\begin{eqnarray}\label{eq9}
		&\hat{H}_{rel} = - \dfrac{\hbar^2}{2\mu_x}\dfrac{\partial^2}{\partial x^2}
		- \dfrac{\hbar^2}{2\mu_y}\dfrac{\partial^2}{\partial y^2} \qquad\qquad\nonumber\\
		&+ \dfrac{e^2B^2}{8}\!\left(
		\dfrac{\beta_y+\alpha^2}{\mu_y}x^2 + \dfrac{\beta_x+\alpha^2}{\mu_x}y^2\right) \nonumber\\
		& \qquad - i\hbar\,\alpha\dfrac{ e B}{2}\!\left(
		\dfrac{y}{\mu_y} \dfrac{\partial}{\partial x} - \dfrac{x}{\mu_x} \dfrac{\partial}{\partial y}
		\right) + V_{RK}(r)\nonumber\\
		&+\dfrac{eB}{M_x}K_yx-\dfrac{eB}{M_y}K_xy,
	\end{eqnarray}
	where the dimensionless coefficients are defined as
	\begin{eqnarray}\label{eq10}
		& \alpha=\dfrac{1-\rho_x\rho_y}{(1+\rho_x) (1+\rho_y)}, 
		\nonumber\\
		& \beta_x=\dfrac{4\rho_x}{(1+\rho_x)^2},\quad
		\beta_y=\dfrac{4\rho_y}{(1+\rho_y)^2},
	\end{eqnarray}
	with mass ratios $\rho_x= m_{x}^e/m_{x}^h$ and $\rho_y= m_{y}^e/m_{y}^h$.
	
	Equation~\eqref{eq9} represents the exact relative-motion Hamiltonian for an anisotropic two-dimensional magnetoexciton. The anisotropy is reflected in both the kinetic terms through the reduced masses $\mu_x$ and $\mu_y$, and the magnetic contributions through the coefficients $\alpha$, $\beta_x$, and $\beta_y$. The presence of the linear derivative term proportional to $\alpha$ indicates the magnetic-field-induced coupling between different angular components of the wave function. The quadratic magnetic terms contain modified coefficients that are properly reduced to unity in the isotropic limit.
	
	Thus, without invoking any stationary-c.m. approximation, we obtain an exact separation of the c.m. motion and a closed Schrödinger equation \eqref{eq9} for the relative motion of the anisotropic magnetoexciton.

	\subsection{The Schr\"{o}dinger for relative motion}
	
	In Eq.~\eqref{eq9}, the relative-motion Hamiltonian contains terms proportional to the pseudomomentum components $K_x$ and $K_y$. These terms describe the coupling between the relative motion and the c.m. motion in the presence of a magnetic field. In general, the exciton c.m. may carry finite momentum determined by thermal excitation or external perturbations. For a thermalized exciton ensemble, the kinetic energy of the c.m. motion can be estimated as $k_B T$, where $k_B$ is the Boltzmann constant and $T$ is the temperature. Using the equipartition theorem
	\begin{equation}
		\frac{1}{2M_x}\,K_x^2\sim k_B T,\quad \frac{1}{2M_y}\,K_y^2\sim k_B T,\nonumber
	\end{equation}
	one obtains characteristic values 
	\begin{equation}
		K_x\sim \sqrt{2M_xk_BT}, \quad  K_y\sim \sqrt{2M_y k_BT}.   \nonumber
	\end{equation}
	
	In typical low-temperature magneto-optical experiments on 2D semiconductors, the exciton c.m. momentum is small \cite{Ly2023b}. Therefore, in the following analysis, we focus on the case $K_x = K_y = 0$, which corresponds to excitons near the bottom of the dispersion and captures the dominant contribution to optical transitions. Under this condition, the Schr\"{o}dinger equation for relative motion becomes
	\begin{eqnarray}\label{eq11}
		&\left\{- \dfrac{1}{2\mu_x}\dfrac{\partial^2}{\partial x^2}
		- \dfrac{1}{2\mu_y}\dfrac{\partial^2}{\partial y^2}
		- i\dfrac{\alpha B}{2}\!\left(
		\dfrac{y}{\mu_x} \dfrac{\partial}{\partial x} -  \dfrac{x}{\mu_y} \dfrac{\partial}{\partial y}
		\right)\right.\nonumber\\
		&\left.+ \dfrac{B^2}{8}\left( \dfrac{\beta_y+\alpha^2}{\mu_y} x^2 + \dfrac{\beta_x+\alpha^2}{\mu_x} y^2
		\right)+ V_{RK}(r)\right\}\psi(x,y) \nonumber\\
		&= E \,\psi(x,y),
	\end{eqnarray}
	where we have expressed the equation in atomic units. In these units, energies are measured in Hartree energy $E_H$, distances in Bohr radius $a_0$, and the magnetic field in units of $B_0=(m_e/e\hbar) E_H$=23 051.7 tesla (T).
	The reduced masses $\mu_x$  and $\mu_y$ are given in units of the electron mass $m_e$.
	
	The interaction potential $V_{RK}(r)$ remains isotropic, as the dielectric screening of the surrounding medium is assumed to be isotropic. For the numerical calculations presented below, we employ the Rytova-Keldysh potential expressed in the Fourier representation,
	\begin{equation}\label{eq12}
		V_{RK}(r)=-\frac{1}{2\pi\kappa}\int_{-\infty}^{+\infty}\int_{-\infty}^{+\infty}{dq_1 dq_2}\,
		\dfrac{e^{i q_1 x+ i q_2 y}}{q(1+\frac{r_0}{\kappa} q)},
	\end{equation}
	where $q=\sqrt{q_1^2+q_2^2}$, $r_0$ is the screening length (in units of Bohr radius $a_0$), and $\kappa$ is the surrounding dielectric constant. 
	
	In contrast to the isotropic case, where alternative representations, such as the Laplace transform given in \cite{Ly2023}, may be convenient, the Fourier representation \eqref{eq12} is particularly suitable for anisotropic systems because it preserves the explicit dependence on Cartesian coordinates $x$ and $y$. Equation~\eqref{eq11} therefore provides a starting point for the numerical solution of the anisotropic magnetoexciton problem.
	
	\subsection{Comparison with previous treatments}
	
	The Schrödinger equation for relative motion derived in Eq.~\eqref{eq11} differs from those obtained in earlier studies~\cite{Kezerashvili2022, Kezerashvili2022b}  employing approximate c.m. separation procedures. In the present formulation, additional anisotropy-dependent terms appear explicitly in the relative-motion Hamiltonian.

	First, Eq.~\eqref{eq11} contains the Zeeman term 
	\begin{equation}
		{\hat H}_{Zeeman}=-\, i\,\frac{\alpha B}{2}
		\!\left(
		\frac{y}{\mu_x} \frac{\partial}{\partial x} - \frac{x}{\mu_y} \frac{\partial}{\partial y}
		\right),
		\label{eq13}
	\end{equation}
	which is absent in treatments based on factorized wave functions for the s states. This term \eqref{eq13} is associated with the orbital motion of the electron-hole pair and is correctly reduced to the known isotropic limit~\cite{Ly2023} when $\mu_x=\mu_y =\mu$ and $\rho_x=\rho_y=\rho$. In this limit, $\alpha=(1-\rho)/(1+\rho)$, and the term becomes a coupling between the magnetic field and the angular momentum. In anisotropic systems, where angular momentum is no longer conserved, this contribution reflects the mixing of different angular components induced by the magnetic field and effective-mass anisotropy. In the calculations presented here, this term is retained for all states.
	
	Second, the diamagnetic term 
	\begin{equation}\label{eq14}
		{\hat H}_{diamag} =\frac{e^2 B^2}{8}\left(\frac{\beta_y+\alpha^2}{\mu_y}\,x^2 
		+ \frac{\beta_x+\alpha^2}{\mu_x}\,y^2\right)\quad
	\end{equation}
	contains coefficients $\beta_x+\alpha^2 $ and $\beta_y+\alpha^2$ determined by the exact transformation. In the isotropic limit, these coefficients reduce to unity, recovering the standard form of the magnetic confinement term, presented in \cite{Ly2023}. In anisotropic materials, however, their values depend explicitly on the mass ratios $\rho_x$ and $\rho_y$ by Eqs.~\eqref{eq10}. The commonly used approximation, based on the assumption that the wave function factorizes as $\Psi(\mathbf{r}, \mathbf{R})=\exp\,( \frac{i}{\hbar} \mathbf{K}\cdot \mathbf{R}) \psi(\mathbf{r})$ as suggested in Ref.~\cite{Landau1977}, corresponds effectively to the limit of a much heavier hole mass, where $\rho_x, \rho_y \longrightarrow 0$, leading to $\beta_x+\alpha^2=1$ and $\beta_y+\alpha^2=1$.
	
	For realistic material parameters of BP and TiS$_3$ presented in Table~\ref{Tab1}, the coefficients deviate from unity, as shown in Table~\ref{Tab2}. These deviations reflect the intrinsic mass anisotropy and directly influence the magnetic contribution to the exciton energy and the calculated diamagnetic coefficients. Therefore, retaining the exact coefficients ensures internal consistency of the Hamiltonian and provides a quantitatively reliable description of anisotropic magnetoexcitons.
	
	\begin{table}[t]
		\centering
		\caption{\label{Tab1}
			Electron and hole effective masses $m_x^e$, $m_y^e$, $m_x^h$, and $m_y^h$ ($m_e$) in monolayer black phosphorus (BP) and titanium trisulfide (TiS$_3$) from Refs.~\cite{Paez2016, Torun2018, Donck2018}, together with the calculated
			exciton reduced masses $\mu_x$ and $\mu_y$  ($m_e$).
		}
		\setlength{\tabcolsep}{7.2pt}
		\begin{tabular}{lcccccc}
			\toprule
			& $m_x^{e}$ & $m_y^{e}$ & $m_x^{h}$ & $m_y^{h}$ & $\mu_x$ & $\mu_y$ \\
			\midrule
			BP     & 0.199 & 0.753 & 0.168 & 5.353 & 0.091 & 0.660 \\
			TiS$_3$   & 1.52 & 0.40 & 0.30 & 0.99 & 0.251  & 0.285 \\
			\bottomrule
		\end{tabular}
	\end{table}

	\begin{table}[t]
		\centering
		\caption{\label{Tab2}
			Coefficients $\alpha$, $\beta_x+\alpha^2$, and $\beta_y+\alpha^2$ in Eqs.~\eqref{eq13} and \eqref{eq14} calculated using the exact theoretical formulas \eqref{eq10} for BP and TiS$_3$.
		}
		\setlength{\tabcolsep}{8pt}
		\begin{tabular}{lccccc}
			\toprule
			& $\rho_x$ & $\rho_y$ & $\alpha$
			& $\beta_x+\alpha^2$ & $\beta_y+\alpha^2$ \\
			\midrule
			{{BP}}         & 1.176 & 0.140 & 0.337    & 1.107 & 0.545 \\
			{{TiS$_3$}}             & 5.067 & 0.404 & $-0.123$ & 0.566 & 0.835 \\
			\bottomrule
		\end{tabular}
	\end{table}
	
	\begin{figure*} [bt]
		
		\includegraphics[scale=0.26]{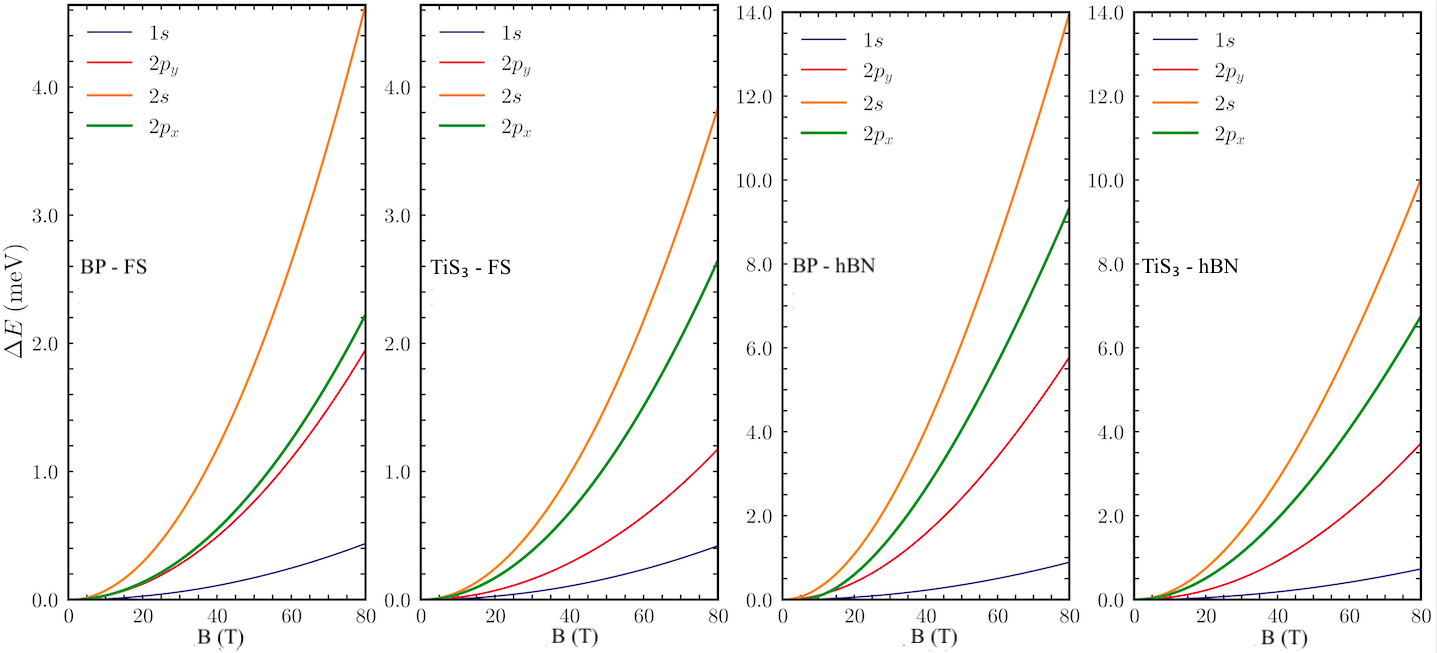}
		\caption{(Color online) Magnetic contribution to the exciton binding energy as a function of the magnetic field $B$: $\Delta E= E(B)- E(B=0)$ for the states 1s, 2p$_y$, 2p$_x$, 2s with r$_0$=2.576 nm (for BP),  r$_0$=4.434 nm (for TiS$_3$), $\kappa = 1.0$ (for FS), and  $\kappa = 4.9$ (for hBN). }\label{Fig1}
	\end{figure*} 
	
	\section{Numerical results for magnetoexciton energies}\label{sec3}
	
	\subsection{Method of calculations}\label{subsecIIIa}
	To solve the relative-motion Schr\"{o}dinger equation \eqref{eq11}, we employ the FK operator method \cite{Feranchuk1982, Feranchuk2015} combined with the application of the Levi-Civita transformation \cite{Hoang1993}. This approach provides a non-perturbative and algebraic framework suitable for anisotropic systems and allows systematically convergent numerical solutions with controllable precision.
	
	First, we perform the Levi-Civita transformation 
	\begin{equation}\label{eq15}
		x=u^2-v^2, \;y=2uv\,,
	\end{equation}
	which maps the original two-dimensional problem into a form analogous to a two-dimensional anharmonic oscillator. This transformation is particularly convenient for treating Coulomb-type interactions and has been successfully applied in related excitonic systems \cite{Nguyen2019, Ly2023, Ly2023b, Ly2025}. 
	
	In the transformed ($u, v$) space, the Hamiltonian can be expressed in terms of creation and annihilation operators,
	\begin{eqnarray}
		\hat a(\omega)=\sqrt{\frac{\omega}{2}}
		\left(u+\frac{1}{\omega}\frac{\partial}{\partial u}\right), \; 
		\hat a^+(\omega)=\sqrt{\frac{\omega}{2}}
		\left(u-\frac{1}{\omega}\frac{\partial}{\partial u}\right), \nonumber\\
		\hat b(\omega)=\sqrt{\frac{\omega}{2}}
		\left(v+\frac{1}{\omega}\frac{\partial}{\partial v}\right), \; 
		\hat b^+(\omega)=\sqrt{\frac{\omega}{2}}
		\left(v-\frac{1}{\omega}\frac{\partial}{\partial v}\right)\,,\nonumber
	\end{eqnarray}
	where $\omega$ is a variational parameter that can be adjusted to optimize convergence. The operators satisfy the commutation relation 
	$[\hat a, \,\hat a^+]=1$ and $[\hat b,\, \hat b^+]=1$.
	
	Next, the wave function is expanded in the complete basis set 
	\begin{eqnarray}\label{eq16}
		|jk\rangle=\frac{1}{\sqrt{(j-k)!(j+k)!}}\, 
		{({\hat a}^+)}^{j+k}{({\hat b}^+)}^{j-k}|0(\omega)\rangle\,, 
	\end{eqnarray}
	where the vacuum state is defined by equations ${\hat a} \,|0\rangle=0$ and ${\hat b} \,|0\rangle=0$; quantum number $j$ is a non-negative number $j=0,1,2,\dots$, while $k$ is an integer less than $j$: $k=-j, -j+1, ..., +j$. The eigenfunction is written as 
	\begin{equation}\label{eq17}
		|\psi\rangle= \sum_{j=0}^{+\infty}\sum_{k=-j}^{+j} C_{jk}\,|jk\rangle.
	\end{equation}
	Substituting this expansion into the Schr{\"o}dinger equation reduces the problem to a matrix eigenvalue equation for the coefficients $C_{jk}$. All matrix elements are evaluated analytically using algebraic manipulation of the operator commutation relations, resulting in an explicit finite-dimensional generalized eigenvalue problem.
	
	In practical calculations, the basis set is truncated at $j \leq N_{\mathrm{max}}$. We typically use $N_{\mathrm{max}}=20$, which ensures that the lowest excitonic energies converge to the desired precision. The convergence was verified by increasing 
	$N_{\mathrm{max}}$ and monitoring the stability of the eigenvalues. Variations of the parameter $\omega$ 
	were also examined to confirm that the resulting energies are insensitive to this parameter choice within an appropriate range. These checks guarantee that the obtained solutions are systematically convergent and numerically stable.
	
	The resulting matrix eigenvalue problem is solved using standard linear algebra routines (e.g., the Linear Algebra Package (LAPACK) \cite{Lapack} or equivalent numerical libraries). This procedure yields magnetoexciton energies and wave functions with controlled numerical accuracy for the anisotropic systems considered.
	
	\subsection{Labeling the quantum states}\label{subsecIIIb}
	
	In anisotropic systems, the magnetic quantum number is no longer a conserved quantity because rotational symmetry is broken. Nevertheless, it is convenient to retain the conventional notation s, p$_x$, p$_y$, etc., in order to facilitate comparison with the isotropic limit and with previous studies.
	
	In the present work, the labeling of quantum states is performed on the basis of the structure of the eigenvectors obtained from the expansion \eqref{eq17}. After solving the matrix eigenvalue problem, each eigenstate is characterized by a set of coefficients $C_{jk}$. The contribution of each basis state $|jk \rangle$ to a given eigenfunction is determined by the magnitude of the corresponding coefficient. The dominant component provides a natural way to assign approximate quantum numbers $(n,m)$ to the state.
	
	In particular, the states labeled as ``s" correspond to those whose dominant contribution originates from basis states with $k=0$. Similarly, the states p$_x$ and p$_y$ are associated with dominant contributions from components corresponding to $k=+1$ and $k=-1$, respectively. Because of anisotropy and magnetic-field-induced coupling, eigenfunctions \eqref{eq17} generally contain admixtures of several angular components; therefore, labels should be understood as approximate classifications rather than exact quantum numbers.
	
	For highly excited states and for systems with strong anisotropy (large ratio $\beta=\mu_y/\mu_x$), the hybridization between the basis states in \eqref{eq17} becomes more pronounced. In such cases, the magnitudes of several coefficients $C_{jk}$ may become comparable, making the identification less straightforward. To ensure consistent labeling, we track the evolution of eigenvalues by gradually varying the anisotropy parameter from the isotropic limit $\beta =1$ to the physical value of the material under consideration. This procedure allows consistent identification of states across the full anisotropy range. Each state can thus be continuously connected to its counterpart in the isotropic case, where the quantum numbers are well-defined. This procedure allows for a systematic and unambiguous assignment of state labels throughout the calculations.

	\subsection{Magnetoexciton energies}\label{subsecIIIc}
	
	We now present the numerical results for magnetoexciton energies obtained from the exact c.m. separation and the systematically convergent FK operator method described above. Calculations are performed for monolayer BP and TiS$_3$ in both FS and hBN-encapsulated configurations. The effective masses are taken from the references listed in Table~\ref{Tab1}, the screening length values $r_0$ are $2.576$ nm for BP \cite{Rodin2014} and $4.434$  for TiS$_3$ \cite{Torun2018}, and the surrounding dielectric constant $\kappa=1$ for FS environments, while the value $\kappa=4.9$ for hBN-encapsulated slabs~\cite{Laturia2018, Kezerashvili2022}.  
	
	As a consistency check, we first consider the case without a magnetic field ($B=0$). The calculated exciton binding energies agree with previously reported theoretical values \cite{Kezerashvili2022, Donck2018} and are consistent with the experimental data available for freestanding BP \cite{Wang2015, Zhang2018}, as shown in Table~\ref{Tab3}. This agreement supports the reliability of both the material parameters and the numerical implementation.
	
	\begin{table}[H]
		\centering
		\caption{\label{Tab3}
			Calculated exciton energies (meV) in BP and TiS$_3$ compared with those in other theoretical and experimental studies for the case without magnetic field ($B=0$).}
		
		\setlength{\tabcolsep}{8.2pt}
		\begin{tabular}{lcccc}
			\toprule
			
			& 1s & 2p$_y$ & 2s & 2p$_x$  \\
			\midrule
			
			\multicolumn{5}{l}{\textbf{BP} (freestanding)}\\
			Our study & 745.73 & 478.56 & 377.57 & 310.93  \\
			Theor. \cite{Kezerashvili2022}  & 746.06 & 478.75 & 377.72 & 311.22  \\
			Exp. \cite{Zhang2018}   & 762 & -- & -- & --  \\
			Exp. \cite{Wang2015}  & $900 \pm 120$ & -- & -- & --  \\
			
			\midrule
			(in hBN)\\
			Our study &  199.01 & 74.06 & 50.13 &  31.83 \\
			Theor. \cite{Kezerashvili2022}  & 199.66 & 74.37 & 50.35 & 32.03  \\
			
			\midrule
			\multicolumn{5}{l}{{\bfseries TiS$_3$} (freestanding)}\\
			Our study & 537.45 & 314.24 & 306.85 & 255.64  \\
			Theor. \cite{Donck2018}  & 537.1 & -- & -- & --  \\
			
			\bottomrule
		\end{tabular}
	\end{table}

	The primary focus of the present work is the magnetic-field dependence of exciton energies. Magnetoexciton energies are computed over a broad magnetic-field range up to 120 T for the ten lowest excitonic states. The complete numerical values are provided in Tables \ref{TabA1}--\ref{TabA4} of the Appendix~\ref{AppA}, offering a comprehensive dataset for anisotropic magnetoexciton spectra.
	
	Figure \ref{Fig1} illustrates the magnetic contribution to the binding energy $\Delta E (B)= E(B) - E(B=0)$ for the four lowest states (1s, 2p$_y$, 2s, and 2p$_x$). The magnetic-field dependence reflects the interplay between anisotropic effective masses and magnetic confinement. In strongly anisotropic materials such as BP, the deviations from isotropic behavior become more pronounced, particularly for excited states. In contrast, for TiS$_3$, where the mass anisotropy is weaker, the magnetic response exhibits a comparatively smoother evolution.
	
	The diamagnetic coefficients are extracted from the low-field quadratic behavior,
	\begin{equation}
		\sigma_{\mathrm{diamag}}=\frac{1}{2}\lim_{B\rightarrow 0}\frac{\partial^2 E(B)}{{\partial B}^2} ,  
	\end{equation}
	and are summarized in Table~\ref{Tab4}. The observed differences between the present values and those reported in Refs.~\cite{Kezerashvili2022, Kezerashvili2022b} are likely related to differences in the treatment of c.m. separation and magnetic coupling terms in anisotropic systems.
	The results show that anisotropy-dependent coupling terms significantly influence the diamagnetic response, especially in BP. Since these coefficients determine the curvature of the energy shift at low magnetic fields, their accurate evaluation is essential for the quantitative interpretation of magneto-optical measurements.
	
	\begin{table}[t]
		\centering
		\caption{\label{Tab4}
			Diamagnetic coefficients in BP and TiS$_3$ monolayers,
			calculated by our method with the exact c.m. separation compared with those of Refs.~\cite{Kezerashvili2022, Kezerashvili2022b}.
		}
		\setlength{\tabcolsep}{4pt}  
		\begin{tabular}{lcccccccc}
			\toprule
			& \multicolumn{4}{c}{\textbf{BP}} & \multicolumn{4}{c}{\textbf{TiS$_3$}} \\
			\cmidrule{2-5} \cmidrule{6-9}
			\textbf{State}
			& \multicolumn{2}{c}{Our work}
			& \multicolumn{2}{c}{Ref.~\cite{Kezerashvili2022}}
			& \multicolumn{2}{c}{Our work}
			& \multicolumn{2}{c}{Ref.~\cite{Kezerashvili2022b}} \\
			\cmidrule{2-3} \cmidrule{4-5}
			\cmidrule{6-7} \cmidrule{8-9}
			& hBN & FS & hBN & FS & hBN & FS & hBN & FS \\
			\midrule
			1 & 0.14 & 0.07 & 0.46 & 0.21 & 0.11 & 0.07 & 0.17 & 0.095 \\
			2 & 1.02 & 0.31 & 1.41 & 0.40 & 0.56 & 0.18 & --- & 0.04 \\
			3 & 2.70 & 0.74 & 2.96 & 0.67 & 1.27 & 0.43 & --- & 0.46 \\
			4 & ---  & 0.34 & ---  & 1.88 & 1.87 & 0.61 & --- & 0.88 \\
			\bottomrule
		\end{tabular}
	\end{table}
	
	The availability of systematic numerical data for ten excitonic states, presented in Tables~\ref{TabA1}--\ref{TabA4}, enables detailed analysis of higher excited states, whose magnetic evolution is sensitive to both effective-mass anisotropy and screening effects. These results provide a consistent and reproducible reference for future theoretical modeling and comparison with high-field magneto-optical experiments.
	
	\subsection{Wave functions}\label{SubsecIIID}
	
	\begin{figure*}[!htbp]
		\centering
		\includegraphics[scale=0.25]{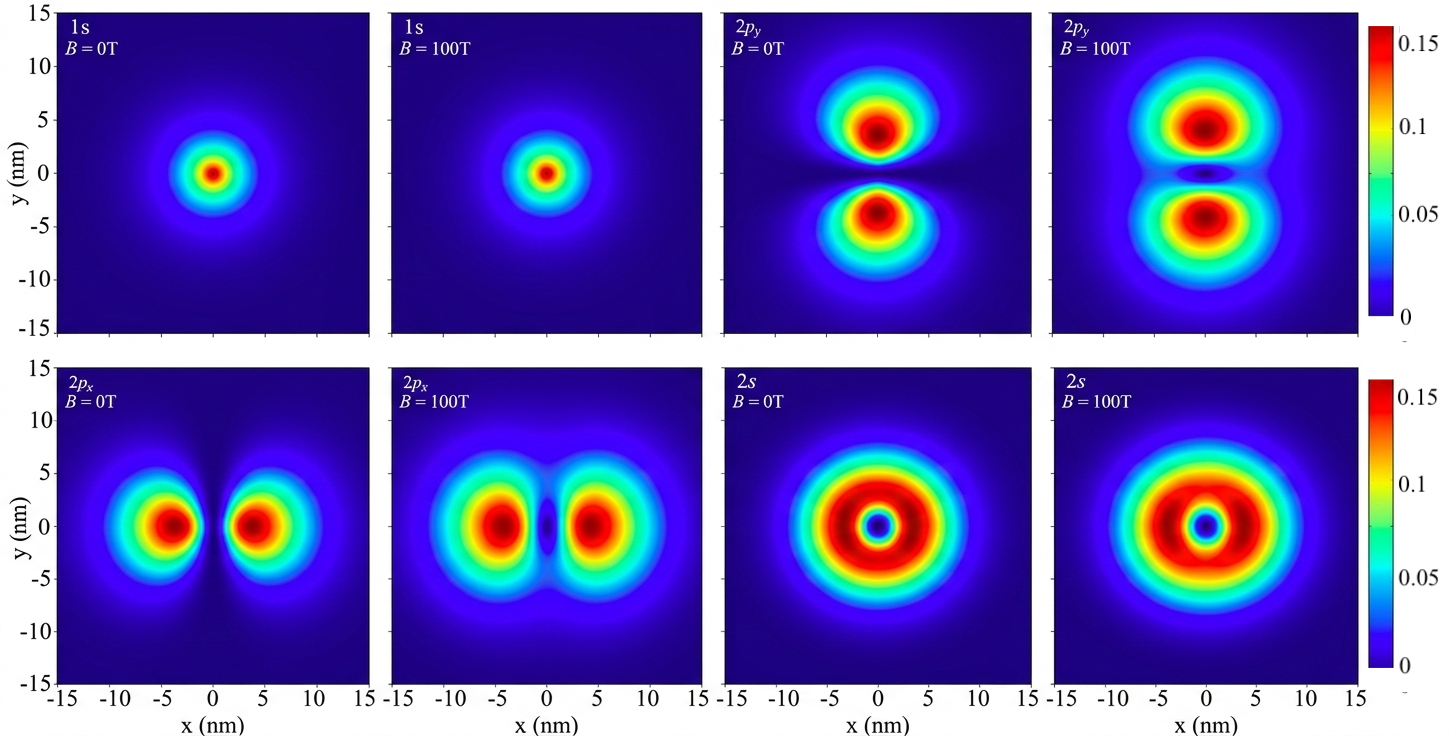}
		\caption{(Color online) The probability density for a magnetoexciton in TiS$_3$.}
		\label{Fig2}
	\end{figure*}

	\begin{figure*}[!htbp]
		\centering
		\includegraphics[scale=0.25]{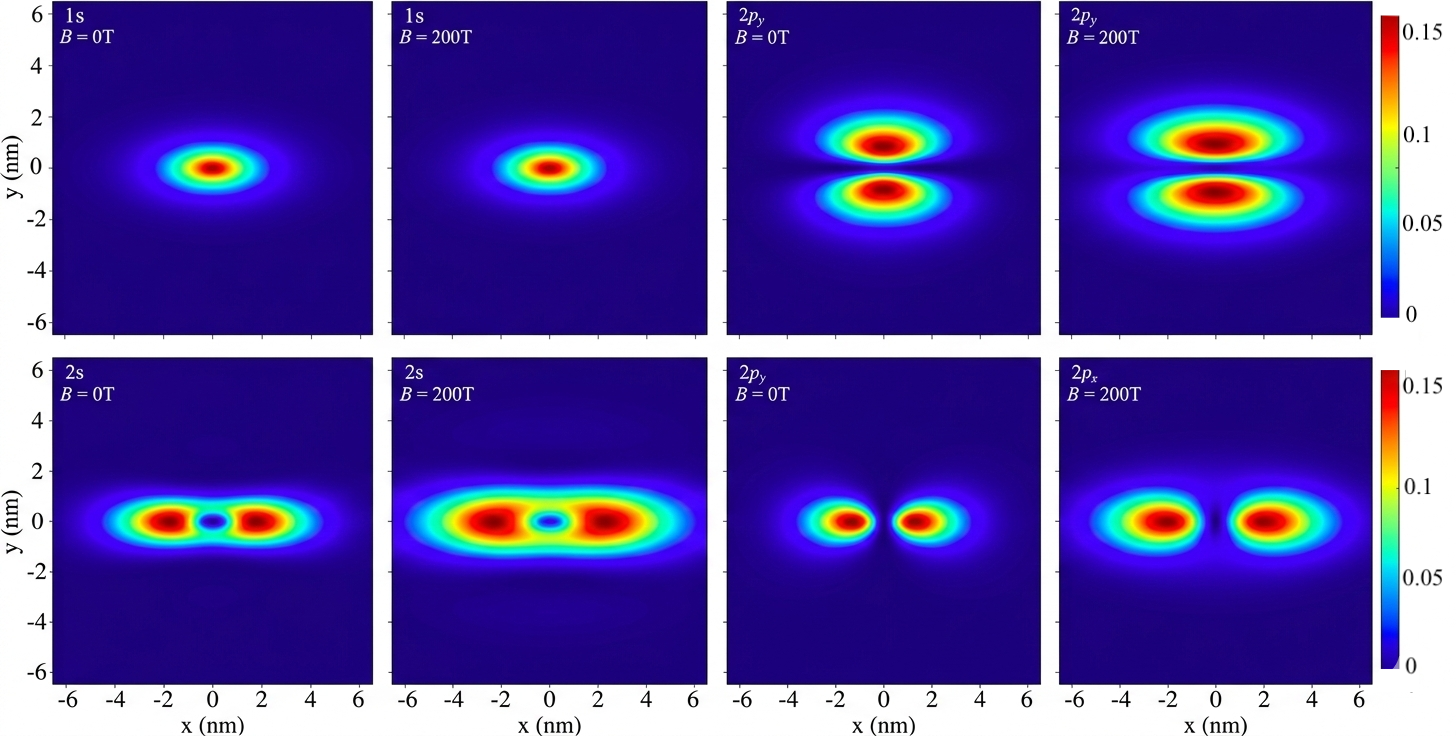}
		\caption{(Color online) The probability density for a magnetoexciton in BP.}
		\label{Fig3}
	\end{figure*}

	To obtain a more comprehensive understanding of how the surrounding environment influences the exciton's spatial distribution, we compute wave functions and visualize the corresponding probability densities for the following scenarios. First, we consider the anisotropy case, keeping the magnetic field fixed at $B = 0$. Second, we include both anisotropy and a finite magnetic field. The results are shown for TiS$_3$ for the four excitonic states 1s, 2s, 2p$_x$, and 2p$_y$. The purpose of this analysis is to examine how the exciton probability density evolves across different environments, starting from the case with only anisotropy and progressing to the case with both anisotropy and a magnetic field. We can see that, although the anisotropy of TiS$_3$ is relatively low, its influence on the electron distributions is significant, as shown in Fig.~\ref{Fig2} for TiS$_3$ encapsulated by hBN. A similar scenario for BP is presented in Fig.~\ref{Fig3}. 
	
	As mentioned above, we label the first four states by 1s, 2s, 2$p_x$, 2$p_y$, while other theoretical studies label them as 1s, 2s, 3s, and 4s \cite{Donck2018, Kezerashvili2022, Kezerashvili2022b}. For genuine quasi-s states, the wave function must satisfy $|\psi(0)|^2 \neq 0$. In contrast, for states with an azimuthal quantum number $m \neq 0$, we have $|\psi(0)|^2 = 0$. The electron distributions given in Figs.~\ref{Fig2} and \ref{Fig3} demonstrate that our definitions are more appropriate.

	\section{Conclusion}\label{concl}
	
	In this work, we have derived an exact analytical procedure for separating the center-of-mass (c.m.) and relative motions of an anisotropic two-dimensional exciton in a perpendicular magnetic field. By employing the conserved pseudomomentum, we obtained a relative-motion Hamiltonian that explicitly contains anisotropy-dependent coupling terms and modified magnetic coefficients arising from direction-dependent effective masses. These terms influence the magnetic contribution to the exciton binding energy and the associated diamagnetic response.
	
	The resulting Schrödinger equation was solved using the Feranchuk–Komarov operator method combined with the Levi-Civita transformation. This non-perturbative approach provides systematically convergent numerical solutions with controllable precision, enabling accurate determination of magnetoexciton energies and wave functions in anisotropic systems.
	
	Applying the formalism to monolayer black phosphorus and titanium trisulfide in both freestanding and hBN-encapsulated configurations, we calculated magnetoexciton energies, diamagnetic coefficients, and probability densities for the ten lowest excitonic states over a broad range of magnetic fields. The comprehensive numerical data tables presented in this work provide a consistent and reproducible reference for anisotropic magnetoexciton spectra and may serve as benchmarks for future theoretical and magneto-optical investigations.
	
	The formalism developed here is general and can be extended to other anisotropic 2D semiconductors and heterostructures. Possible directions for further study include finite pseudomomentum effects at elevated temperatures, excitonic complexes such as trions and biexcitons, and strain-induced anisotropy. The present results contribute to a systematic theoretical framework for magnetoexcitons in low-symmetry two-dimensional materials.
	
	\section*{Acknowledgments} This research is funded by Vietnam Ministry of Education and Training, Grant number B2025-SPS-03 and carried out by the high-performance cluster at Ho Chi Minh City University of Education, Vietnam.
	
	Contribution: V.-H.L. and D.-K.D.L. conceptualized the work, developed the methodology, and performed analytical formulation; D.-K.D.L. and H.-V.L. carried out numerical calculations; V.-H.L., D.-K.D.L., and D.-N.Le analyzed the data and contributed to the interpretation of the physical underlying; H.-V.L., D.-A.P.N., T.-S.N., and N.-T.D.H. validated the data; V.-H.L., D.-K.D.L., and D.-N.L. wrote the original draft; N.-T.D.H. acquired the funding; V.-H.L. and N.-T.D.H. supervised the work. All authors discussed the results and contributed to the review, editing, and finalization of the manuscript.
	
	\onecolumngrid
	\appendix
	\section{Magnetoexciton energies}\label{AppA}
	
	For completeness and reproducibility, we list the calculated magnetoexciton energies for the ten lowest states at magnetic field intensities up to 120 T for monolayer black phosphorus (BP) and titanium trisulfide (TiS$_3$), considering both freestanding and hexagonal boron nitride (hBN)-encapsulated environments.

	\begin{table}[H]
		\caption{Magnetoexciton energies in monolayer BP in FS with $r_0 = 2.576\,\mathrm{nm}$,
			$\mu_x = 0.091\,m_e$, $\kappa = 1.0$ and $\beta=7.249$.}\label{TabA1}
		\footnotesize
		\vspace{0.2cm}
		\begin{ruledtabular}
			\begin{tabular}{c r r r r r r r r r r}
				
				Magnetic field   &   \multicolumn{10}{c} {Energy (meV)} \\
				(T) & 1s & 2p$_y$ & 2s & 2p$_x$  & 3p$_y$ & 3s & 3d$_y$ & 4s & 3p$_x$ & 4p$_y$ \\
				\hline
				&&&&&&&&&&\\
				0  & -745.738 & -478.564 & -377.573 & -310.933 & -297.502 & -250.417 & -233.884 & -207.986 & -196.664 & -192.115 \\
				2  & -745.738 & -478.564 & -377.573 & -310.933 & -297.502 & -250.417 & -233.884 & -207.986 & -196.664 & -192.115 \\
				4  & -745.737 & -478.560 & -377.565 & -310.929 & -297.484 & -250.390 & -233.872 & -207.940 & -196.647 & -192.089 \\
				6  & -745.736 & -478.554 & -377.550 & -310.922 & -297.452 & -250.344 & -233.852 & -207.863 & -196.618 & -192.046 \\
				8  & -745.734 & -478.545 & -377.529 & -310.913 & -297.409 & -250.281 & -233.823 & -207.756 & -196.578 & -191.985 \\
				10 & -745.732 & -478.534 & -377.502 & -310.901 & -297.353 & -250.200 & -233.787 & -207.618 & -196.527 & -191.907 \\
				20 & -745.711 & -478.442 & -377.281 & -310.798 & -296.888 & -249.530 & -233.482 & -206.481 & -196.106 & -191.257 \\
				30 & -745.677 & -478.288 & -376.912 & -310.627 & -296.118 & -248.435 & -232.967 & -204.631 & -195.426 & -190.175 \\
				40 & -745.629 & -478.074 & -376.398 & -310.386 & -295.052 & -246.950 & -232.231 & -202.135 & -194.508 & -188.660 \\
				60 & -745.492 & -477.463 & -374.943 & -309.690 & -292.070 & -243.028 & -229.990 & -195.612 & -192.025 & -184.254 \\
				70 & -745.403 & -477.068 & -374.006 & -309.235 & -290.179 & -240.750 & -228.394 & -191.864 & -190.482 & -181.255 \\
				80 & -745.300 & -476.614 & -372.935 & -308.710 & -288.038 & -238.400 & -226.389 & -188.747 & -188.025 & -177.606 \\
				90 & -745.184 & -476.103 & -371.732 & -308.115 & -285.660 & -236.088 & -223.898 & -186.827 & -184.247 & -173.231 \\
				100 & -745.054 & -475.535 & -370.401 & -307.450 & -283.057 & -233.893 & -220.875 & -184.728 & -180.595 & -168.138 \\
				120 & -744.755 & -474.231 & -367.372 & -305.921 & -277.227 & -229.909 & -213.307 & -180.014 & -173.553 & -156.166 \\
				
			\end{tabular}
		\end{ruledtabular}
	\end{table}
	
	\begin{table}[H]
		\caption{Magnetoexciton energies in monolayer BP in hBN with $r_0 = 2.576\,\mathrm{nm}$,
			$\mu_x = 0.091\,m_e$, $\kappa = 4.9$ and $\beta=7.249$.}\label{TabA1}
		\footnotesize
		\vspace{0.2cm}
		\begin{ruledtabular}
			\begin{tabular}{c r r r r r r r r r r}
				
				Magnetic field   &   \multicolumn{10}{c} {Energy (meV)} \\
				(T) & 1s & 2p$_y$ & 2s & 2p$_x$  & 3p$_y$ & 3s & 3d$_y$ & 4s & 3p$_x$ & 4p$_y$ \\
				\hline
				&&&&&&&&&&\\
				0    &  -199.011 & -74.065 & -50.139 & -31.836 & -31.419 & -23.921 & -17.974 & -14.356 & -12.691 & -12.425 \\
				2    & -199.011 & -74.060 & -50.128 & -31.840 & -31.377 & -23.882 & -17.954 & -14.320 & -12.703 & -12.323 \\
				4    & -199.009 & -74.048 & -50.095 & -31.842 & -31.261 & -23.765 & -17.894 & -14.212 & -12.682 & -12.075 \\
				6    & -199.006 & -74.028 & -50.039 & -31.832 & -31.082 & -23.573 & -17.794 & -14.035 & -12.588 & -11.721 \\
				8    & -199.002 & -73.999 & -49.962 & -31.803 & -30.848 & -23.308 & -17.651 & -13.790 & -12.415 & -11.270 \\
				10   & -198.997 & -73.962 & -49.863 & -31.755 & -30.562 & -22.973 & -17.466 & -13.479 & -12.163 & -10.724 \\
				20   & -198.955 & -73.658 & -49.057 & -31.240 & -28.419 & -20.523 & -15.792 & -11.069 & -9.839  & -6.625 \\
				30   & -198.885 & -73.162 & -47.772 & -30.356 & -25.277 & -17.748 & -12.397 & -7.590  & -6.333  & -3.179 \\
				40   & -198.787 & -72.487 & -46.072 & -29.186 & -21.403 & -15.385 & -7.231  & -3.445  & -2.610  & 0.915 \\
				60   & -198.508 & -70.657 & -41.652 & -26.182 & -12.184 & -10.899 & 4.668    & 5.366    & 6.412    & 11.546 \\
				70   & -198.328 & -69.533 & -39.025 & -24.416 & -8.536  & -7.058  & 8.423    & 10.843  & 13.262  & 17.382 \\
				80   & -198.121 & -68.288 & -36.170 & -22.505 & -6.071  & -1.684  & 12.288  & 16.175  & 20.741  & 23.428 \\
				90   & -197.888 & -66.934 & -33.114 & -20.467 & -3.510  & 3.888    & 16.255  & 21.627  & 28.437  & 29.642 \\
				100 & -197.629 & -65.483 & -29.883 & -18.316 & -0.863  & 9.621    & 20.316  & 27.204  & 35.986  & 36.001 \\
				120 & -197.034 & -62.324 & -22.968 & -13.716 & 4.663    & 21.444  & 28.691  & 38.679  & 48.129  & 49.080 \\
				
			\end{tabular}
		\end{ruledtabular}
	\end{table}
	
	\begin{table}[H]
		\caption{Magnetoexciton energies in monolayer TiS$_3$ in FS with $r_0=4.434$ nm, $\mu_x=0.251\, m_e$, $\kappa=1.0$ and $\beta=1.135$.}\label{TabA1}
		\footnotesize
		\vspace{0.2cm}
		\begin{ruledtabular}
			\begin{tabular}{c r r r r r r r r r r}
				
				Magnetic field   &   \multicolumn{10}{c} {Energy (meV)} \\
				(T) & 1s & 2p$_y$ & 2s & 2p$_x$  & 3p$_y$ & 3s & 3d$_y$ & 4s & 3p$_x$ & 4p$_y$ \\
				\hline
				&&&&&&&&&&\\
				0  & -537.456 & -314.242 & -306.859 & -255.647 & -208.980 & -208.655 & -187.563 & -181.891 & -159.312 & -150.324 \\
				2    & -537.456 & -314.242 & -306.859 & -255.647 & -208.980 & -208.655 & -187.563 & -181.892 & -159.312 & -150.324 \\
				4    & -537.455 & -314.240 & -306.854 & -255.640 & -209.053 & -208.563 & -187.550 & -181.872 & -159.287 & -150.464 \\
				6    & -537.454 & -314.236 & -306.846 & -255.627 & -209.133 & -208.451 & -187.528 & -181.839 & -159.246 & -150.589 \\
				8    & -537.452 & -314.231 & -306.834 & -255.610 & -209.211 & -208.328 & -187.497 & -181.793 & -159.188 & -150.702 \\
				10   & -537.450 & -314.225 & -306.818 & -255.588 & -209.285 & -208.197 & -187.457 & -181.734 & -159.114 & -150.802 \\
				20   & -537.430 & -314.172 & -306.690 & -255.404 & -209.569 & -207.435 & -187.126 & -181.244 & -158.500 & -151.102 \\
				30   & -537.397 & -314.083 & -306.477 & -255.099 & -209.700 & -206.510 & -186.574 & -180.437 & -157.488 & -151.082 \\
				40   & -537.351 & -313.957 & -306.182 & -254.674 & -209.678 & -205.431 & -185.805 & -179.325 & -156.094 & -150.754 \\
				60   & -537.220 & -313.592 & -305.350 & -253.468 & -209.183 & -202.821 & -183.627 & -176.249 & -152.250 & -149.238 \\
				80   & -537.036 & -313.067 & -304.212 & -251.808 & -208.121 & -199.643 & -180.640 & -172.149 & -147.149 & -146.695 \\
				90   & -536.925 & -312.742 & -303.536 & -250.813 & -207.389 & -197.855 & -178.864 & -169.757 & -145.081 & -144.186 \\
				100 & -536.801 & -312.375 & -302.791 & -249.713 & -206.531 & -195.941 & -176.910 & -167.157 & -143.262 & -140.978 \\
				120 & -536.513 & -311.513 & -301.104 & -247.212 & -204.457 & -191.758 & -172.512 & -161.392 & -139.060 & -133.899 \\
				
			\end{tabular}
		\end{ruledtabular}
	\end{table}

	\begin{table}[H]
		\caption{Magnetoexciton energies in monolayer TiS$_3$ in FS with $r_0=4.434$ nm, $\mu_x=0.251\, m_e$, $\kappa=4.9$ and $\beta=1.135$.}\label{TabA4}
		\footnotesize
		\vspace{0.2cm}
		\begin{ruledtabular}
			\begin{tabular}{c r r r r r r r r r r}
				
				Magnetic field   &   \multicolumn{10}{c} {Energy (meV)} \\
				(T) & 1s & 2p$_y$ & 2s & 2p$_x$  & 3p$_y$ & 3s & 3d$_y$ & 4s & 3p$_x$ & 4p$_y$ \\
				\hline
				&&&&&&&&&&\\
				0    & -172.543 & -54.571 & -51.964 & -41.138 & -22.900 & -22.882 & -21.784 & -20.605 & -17.687 & -11.562 \\
				2    & -172.542 & -54.569 & -51.958 & -41.130 & -22.982 & -22.769 & -21.764 & -20.575 & -17.652 & -11.685 \\
				4    & -172.541 & -54.562 & -51.943 & -41.108 & -23.041 & -22.615 & -21.705 & -20.485 & -17.548 & -11.738 \\
				6    & -172.539 & -54.551 & -51.917 & -41.070 & -23.069 & -22.431 & -21.607 & -20.336 & -17.377 & -11.721 \\
				8    & -172.535 & -54.536 & -51.880 & -41.017 & -23.066 & -22.217 & -21.469 & -20.130 & -17.139 & -11.636 \\
				10   & -172.531 & -54.516 & -51.834 & -40.949 & -23.035 & -21.973 & -21.292 & -19.870 & -16.836 & -11.485 \\
				20   & -172.497 & -54.348 & -51.454 & -40.391 & -22.480 & -20.357 & -19.868 & -17.875 & -14.469 & -9.876 \\
				30   & -172.439 & -54.061 & -50.845 & -39.494 & -21.379 & -18.197 & -17.688 & -14.984 & -10.993 & -7.223 \\
				40   & -172.359 & -53.651 & -50.034 & -38.293 & -19.865 & -15.626 & -14.948 & -11.475 & -6.785 & -3.961 \\
				60   & -172.130 & -52.469 & -47.901 & -35.120 & -15.951 & -9.601  & -8.310  & -3.219  & 3.007  & 3.601   \\
				80   & -171.812 & -50.844 & -45.212 & -31.116 & -11.199 & -2.744  & -0.628  & 6.119   & 12.076 & 13.979 \\
				90   & -171.620 & -49.884 & -43.695 & -28.861 & -8.591  & 0.913   & 3.488   & 11.061  & 16.569 & 19.767 \\
				100 & -171.406 & -48.835 & -42.079 & -26.461 & -5.856  & 4.695   & 7.748   & 16.138  & 21.207 & 25.711 \\
				120 & -170.915 & -46.499 & -38.584 & -21.286 & -0.067  & 12.571  & 16.616  & 26.583  & 30.887 & 37.973 \\
				
			\end{tabular}
		\end{ruledtabular}
	\end{table}
	
	\twocolumngrid
	

	\nocite{*}
	\renewcommand{\refname}{References}
	\bibliography{ref}
	

\end{document}